\documentclass[fleqn,10pt]{wlscirep}
\title{The influence of statistical properties of Fourier coefficients on random surfaces}

\author[1,2,*]{C. P. de Castro}
\author[2]{M. Lukovi\'c}
\author[1]{R. F. S. Andrade}
\author[2,3]{H. J. Herrmann}
\affil[1]{Instituto de F\'isica, Universidade Federal da Bahia, Campus Universit\'ario da Federa\c{c}\~ao, Salvador, BA,  40170-115, Brazil}
\affil[2]{ Computational Physics for Engineering Materials, IfB, ETH Zurich, Wolfgang-Pauli-Strasse 27, CH-8093 Zurich, Switzerland}
\affil[3]{Departamento de F\'isica, Universidade Federal do Cear\'a, Fortaleza, Cear\'a,60451-970, Brazil}
\affil[*]{ccastro@ethz.ch}

\begin{abstract}
Many examples of natural systems  can be described by random Gaussian surfaces. 
Much can be learned by analyzing the Fourier expansion of the surfaces, from which it is possible to determine the corresponding Hurst exponent and consequently establish the presence of scale invariance. We show that this symmetry is not affected by the distribution of the modulus of the Fourier coefficients. Furthermore, we investigate the role of the Fourier phases of random surfaces. In particular, we show how the surface is affected by a non-uniform distribution of phases. 
\end{abstract}
\begin{document}

\flushbottom
\maketitle
%
%
 \thispagestyle{empty}

\section*{Introduction}

Two-dimensional random surfaces can be considered as a generalization of one-dimensional stochastic processes.
Often, properties of natural systems, such as sea surface temperatures, rough graphene surfaces and 2D turbulence can be mapped onto random surfaces \cite{Gotsmann2012QuantizedSurfaces,deAssis2012DistributionProfiles,Jiang2016Long-RangeTemperature,Bernard2006ConformalTurbulence,Bernard2007InverseCurves,Giordanelli2016ConformalSheets}. 
Their scaling properties can be characterized by a single parameter known as the Hurst exponent, $H$. 
This exponent is related to the degree of spatial correlation between two points on the surface. 
For all $H > -1$ the surfaces are long-range correlated, rough and self-affine \cite{Giordanelli2016ConformalSheets,Kalda2008StatisticalClusters,Olami1996ScalingSurfaces}. Uncorrelated surfaces correspond to an $H$-value of -1.

Much can be learned about the properties of random surfaces by studying the paths of constant height (lines) extracted from them \cite{Isichenko1992PercolationMedia,Schrenk2013PercolationDisorder,Kalda2008StatisticalClusters,Schmittbuhl1993PercolationSurfaces,Prakash1992StructuralPercolation,Weinrib1984Long-rangePercolation}.
Empirical and numerical studies of these paths suggest that at the height corresponding to the percolation threshold they are scale invariant and that their fractal dimension depends on the Hurst exponent $H$ \cite{Mandelbrot1982TheNature,Kondev1995GeometricalSurfaces,Schrenk2013PercolationDisorder}. 
In some cases they also have an additional symmetry reflected by the conformal invariance of these paths \cite{Bernard2006ConformalTurbulence,Giordanelli2016ConformalSheets}. 
This means that the statistics of such curves is covariant with respect to local scale transformations \cite{Boffetta2008HowShorelines}.

There exist several methods to generate random surfaces \cite{Barnsley1988TheImages}. 
In this work, we will consider the Fourier Filtering Method (FFM), where one first creates a random surface in the reciprocal space and then Fourier transforms it into real space.

In the context of random surfaces, it is taken for granted that critical exponents, such as the fractal dimension of the percolation cluster and its perimeters, the correlation length and the susceptibility, depend only on $H$ \cite{Dietrich1985IntroductionTheory}. In the case of conformal invariance, the current view is not as straightforward. In particular, curves with well defined Hurst exponents do not necessarily exhibit conformal invariance. For example, Bernard \textit{et al.} observed conformal invariance in the iso-height lines of vorticity fields of 2D turbulence \cite{Bernard2006ConformalTurbulence}. They also showed, however, that this properties is violated for iso-height lines extracted from surfaces with the same Hurst exponent but with randomly distributed phases of the surface variables in Fourier space. Therefore, it seems that it is not only the Hurst exponent that plays a determinant role in conformal invariance.

The possible dependence of conformal invariance on phase correlations \cite{Bernard2006ConformalTurbulence} has therefore motivated us to investigate whether the scale invariance of iso-height lines is also affected in a similar way. Given that each point of the random surface in reciprocal space is determined by the phase, as well as the magnitude of a complex number, for completeness we also study the effects of the latter on the scale invariance of the iso-height lines.  We therefore investigate how the critical exponents are influenced by Fourier phases, especially their correlations, as well as the distribution of the magnitudes of the Fourier components.

We show that a non-uniform distribution of Fourier phases introduces symmetries in random surfaces and that an increase in phase correlations in Fourier space is equivalent to a translation of the surface in real space. Furthermore, our results show that changes in the shape of the distribution of Fourier magnitudes, without altering their correlations have the sole effect of modifying the height magnitudes of the random surfaces. None of the variations described above do significantly change the H-dependence of the critical exponents as conjectured Schrenk K. J. in \cite{Schrenk2013PercolationDisorder}.

\section*{Method}
\subsection*{Gaussian surfaces}

A set of random real numbers may be interpreted as a surface, where each number corresponds to the height $h(\textbf{x}) = h(x_{1},x_{2}) $ at coordinate $x_{1}$ and $x_{2}$ \cite{Weinrib1984Long-rangePercolation,Prakash1992StructuralPercolation,Schrenk2013PercolationDisorder,Barnsley1988TheImages,Schrenk2012FracturingSurfaces}.
In order to create correlated random surfaces, we used the Fourier Filtering Method (FFM)\cite{Makse1996MethodSystems,Lauritsen1993EffectPhenomena,Oliveira2011Optimal-pathLattices,Ballesteros1999Site-dilutedDisorder,Morais2011FractalityLandscapes,Fehr2011ScalingWatersheds}, which consists in defining a complex function $\eta(\textbf{q})$ in Fourier space and then taking the inverse transform to obtain $h(\textbf{x})$. The complex Fourier coefficients $\eta(\textbf{q})$ can be written in the form
\begin{equation}
\label{coefficient}
\eta(\textbf{q}) = \texttt{c}(\textbf{q})  \exp(2\pi\phi(\textbf{q})),
\end{equation}
where $\textbf{q}=(q_1,q_2)$ is the frequency in Fourier space, $\texttt{c}(\textbf{q})$ the magnitude and $\phi(\textbf{q})$ the phase. In order to obtain a random surface with the desired properties, we chose the power spectrum $S(\textbf{q})$ of the surface in the form of a power law such that
\begin{equation}
\label{power}
S(\textbf{q}) \sim |\textbf{q}|^{-\beta_{c}} = \left( \sqrt{q_{1}^{2}+q_{2}^{2}} \right)^{-\beta_{c}}
\end{equation}
where $\beta_{c} = 2(H+1)$ \cite{Barnsley1988TheImages} defines the Hurst exponent. 
Then, we apply the power-law filter to a real random variable $\textit{u}(\textbf{q})$ obtaining for the magnitude
\begin{equation}
\texttt{c}(\textbf{q}) = [S(\textbf{q})]^{1/2} \textit{u}(\textbf{q}).
\end{equation}
In general, $\textit{u}(\textbf{q})$ is Gaussian distributed and $ \phi(\textbf{q}) \in [0,1]$ is a uniformly distributed noise and $\texttt{c}(\textbf{q})$ must satisfy the conjugate symmetry condition,  $\texttt{c}(\textbf{-q})=  \overline{\texttt{c}(\textbf{q})} $ \cite{Barnsley1988TheImages}. 

The choice of the power spectrum as a filter is justified by the Wiener-Khintchine theorem \cite{Barnsley1988TheImages,D.K.C.MacDonald2006NoiseIntroduction}, which states that the autocorrelation function, $C(\textbf{r})$, of a time series is the Fourier transform of its power spectrum.
Therefore, from the inverse discrete Fourier transform of $\eta(\textbf{q})$ we obtain $h(x_{1},x_{2})$  
\begin{equation}
\label{fourierTransform}
h(x_{1},x_{2}) = \sum_{q_{1}=0}^{N-1} \sum_{q_{2}=0}^{N-1}  \eta_{q_{1}q_{2}} \exp(-2i\pi(q_{1}x_{1}+q_{2}x_{2}))
\end{equation}
 with the desired power-law correlation function \cite{Barnsley1988TheImages,Schrenk2013PercolationDisorder,Kalda2008StatisticalClusters}
\begin{equation}
\label{autocorrelation}
C(\textbf{r})  \sim \textbf{r}^{2H} .
\end{equation}

According to the definition above, if $H=-1$ and therefore $\beta_{c} = 0$, then the power spectrum in eq. \ref{power} becomes independent of the frequency and the surface  uncorrelated (white noise). As $H$ increases from $-1$, height-height correlations are introduced into the surface.

For any random surface defined on a lattice with $H \geq -1$, the percolation threshold $p_{c}$ can be determined using the well established \textit{rank method}. Moreover, recently a conjecture was put forward for the $H$-dependence of the fractal dimension, as well as the other critical exponents, at the corresponding critical point $p_c$ \cite{Schrenk2013PercolationDisorder}. It should also be noted that as a consequence of the extended Harris criterion \cite{Dietrich1985IntroductionTheory,Smirnov2001CriticalPercolation,Schmittbuhl1993PercolationSurfaces,Sandler2004CorrelatedLevel,Weinrib1983CriticalDisorder,Janke2004Harris-LuckLattices}, there are going to be some critical exponents of 2D systems that are not influenced by correlation effects related to $H \in [-1,-3/4]$, implying that for those Hurst values, the exponents are expected to be the same as for the uncorrelated system \cite{Schrenk2013PercolationDisorder}.

In the case of self-affine surfaces, for which $H>0$, the percolation threshold is not well defined, since there is no unique value of the surface height at which the system percolates. 
Nevertheless, in this case also, it is possible to extend some concepts of percolation theory and relate them to $H$ \cite{Kalda2008StatisticalClusters,Olami1996ScalingSurfaces}.

\subsection*{Clusters and perimeters}

\begin{figure}[ht]
\centering
\includegraphics[scale=0.3]{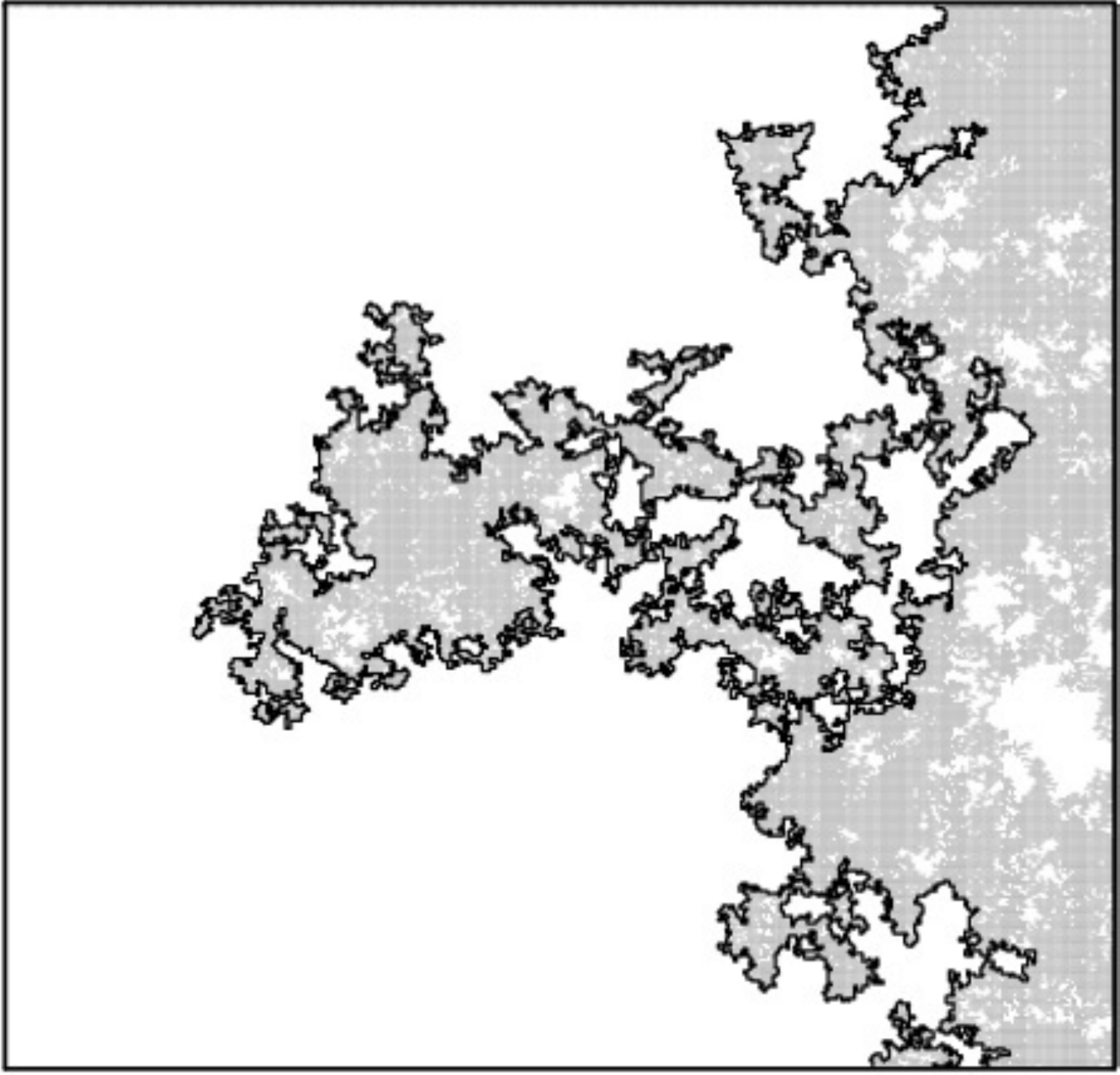}
\caption{Schematic picture of the percolating cluster (light gray) connecting the top of the square with the bottom.
The white region corresponds to sites that do not belong to the percolating cluster (unoccupied sites and other clusters) and the black line is the external perimeter.}
\label{perimeter}
\end{figure}

At the percolation threshold $p_{c}$, occupied neighboring sites create a spanning cluster (percolation cluster) that connects two opposite borders of the surface (fig. \ref{perimeter}). From the percolation cluster we extracted the fractal iso-height lines that correspond to the \textit{complete perimeter} and \textit{accessible perimeters} \cite{Schrenk2013PercolationDisorder,Kalda2008StatisticalClusters,Newman2000EfficientPercolation,Newman2001FastPercolation}. The complete perimeter consists of all bonds between the percolating cluster and unoccupied sites. This is illustrated in fig. \ref{perimeter}, where light grey represents the percolating cluster and the black line follows the complete perimeter.

The accessible perimeter is obtained by eliminating from the complete perimeter all line segments within fjords with a bottleneck equal to the length $r$ of the current stick, according to the yardstick method used to measure the perimeter's fractal dimension. Here, for each value of $r$, the length of any curve is defined by the number of straight yardsticks $N_{r}$ required to go from one extreme to the other by jumping  from one point on the curve to the next at a distance $r$. Then, the fractal dimension $df_{p}$ is defined by
\begin{equation}
N_{r} \sim r^{df_{p}}.
\end{equation}

\begin{figure}[ht]
\centering
\includegraphics[scale=0.6]{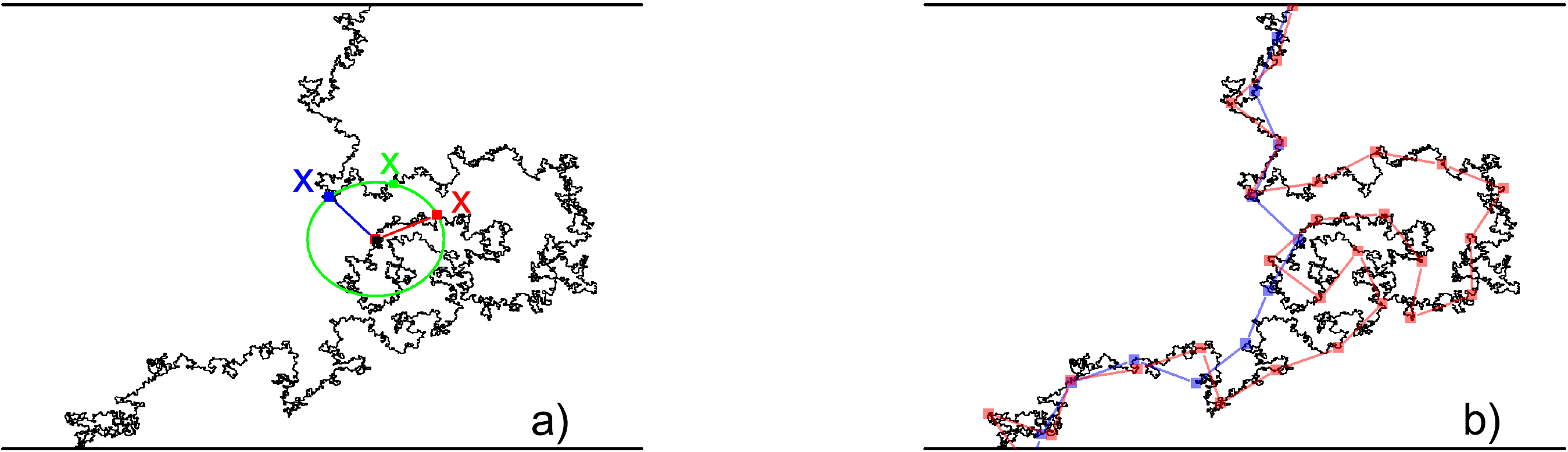}
\caption{ a) Illustration of the rules used to compute the fractal dimension of the complete and accessible perimeters with the yardstick method. Suppose the sticks start to follow the coast from the bottom. The green circle shows the area of coast covered by a particular stick.
The \textbf{X}'s represent the next possible starting points of that particular stick. If the closest point along the coast (red \textcolor{red}{\textbf{X}}) is always chosen as the next starting point, we obtain the complete  perimeter. If, on the other hand, the most distant point (blue \textcolor{blue}{\textbf{X}}) is chosen, then we obtain the accessible perimeter. b) Paths made by sticks of equal sizes of the complete (blue sticks) and accessible (red sticks) perimeters.}
\label{yardstick}
\end{figure}

Fig. \ref{yardstick}.a shows an arbitrary curve where the black dot, in the center of the green circle, indicates the current stick position.
During this specific search for the next point on the curve, three possible positions indicated by red, green and blue X's are found.
If the option to always take the closest position along the curve (red X) is made, the  complete perimeter is obtained. On the other hand, if one always takes the most distant point along the curve (blue dot), which does not avoid the external border, the accessible perimeters is obtained. Indeed, this rule skips points inside fjords and accesses only the external boundary of the coast.
Fig. \ref{yardstick}.b shows the difference between the considered paths, for one specific stick size.

\section*{Results and Discussion}

Having described the method for generating random surfaces using two sets of random variables, $ \textit{u}(\textbf{q})$ and $ \phi(\textbf{q})$, we now discuss how a surface is affected by changing the form of their respective distributions. 

\begin{figure}[ht]
\centering
\includegraphics[scale=0.3]{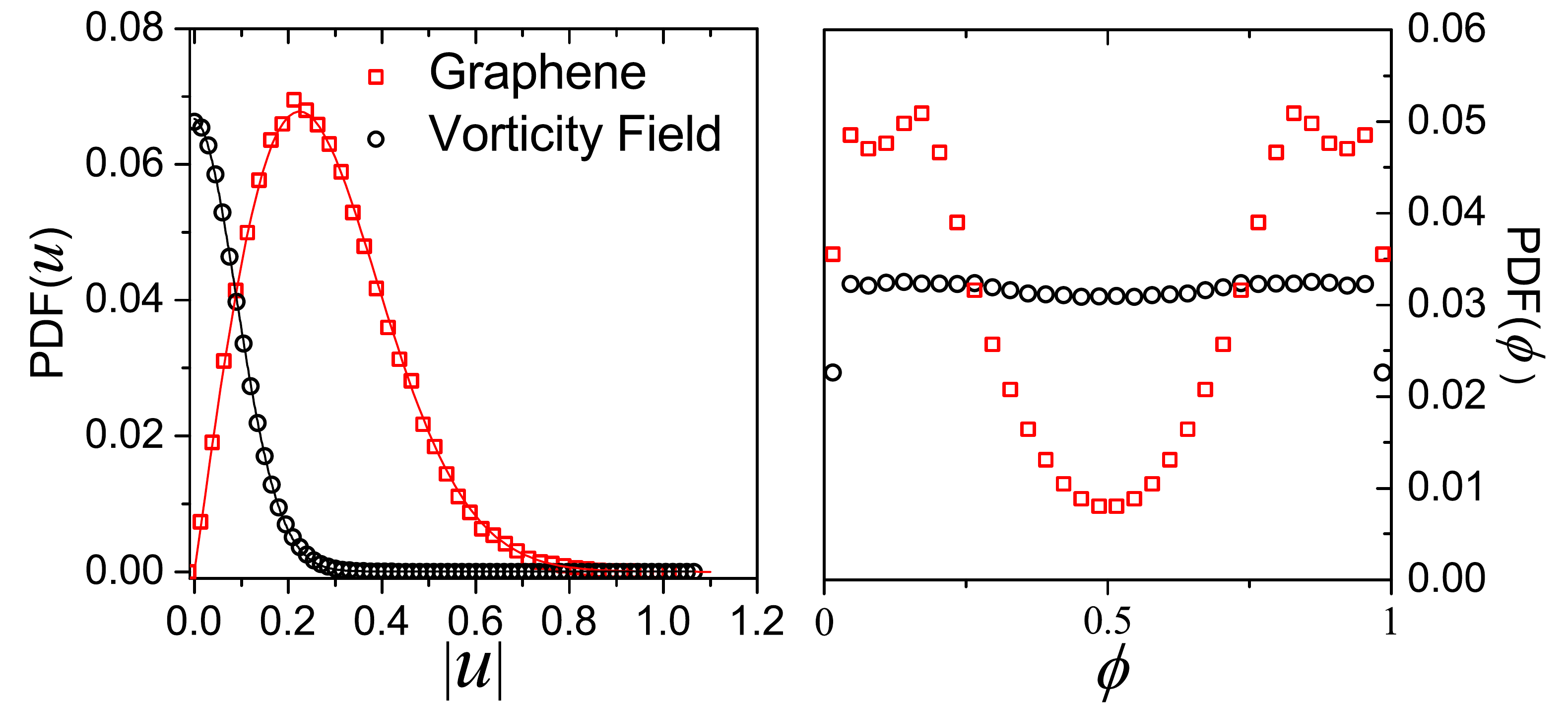}
\caption{Probability density function of $|\textit{u}(\textbf{q})|$  (left) and $ \phi(\textbf{q})$  (right) in the case of the graphene sheet (red squares) and the vorticity field (black circles). The red and black curves in left panel are best fits for $f(|u|) \propto c_{1}|u|\exp^{-c_{2}|u|^{2}}$ and the Gaussian function, respectively.}
\label{distributions}
\end{figure}

Although common \cite{Barnsley1988TheImages}, it is not always the case that $\textit{u}(\textbf{q})$ follows a Gaussian distribution and $ \phi(\textbf{q})$ a uniform one.  For example, Giordanelli \textit{et al.} \cite{Giordanelli2016ConformalSheets} found that for graphene sheets  $ \textit{u}(\textbf{q})$ is well fitted by $f(|u|) \propto c_{1}|u|\exp^{-c_{2}|u|^{2}}$, where $ c_{1},c_{2} $ parameters of the fit. They also found that the Fourier phase distribution $ \phi(\textbf{q})$ is bi-modal and not uniform \cite{Giordanelli2016ConformalSheets}. On the other hand, for the vorticity field of 2D turbulence we confirmed through independent analysis that $ \textit{u}(\textbf{q})$ follows a Gaussian distribution and that the $ \phi(\textbf{q})$ is uniformly distributed. 
Fig. \ref{distributions} compares the distributions of $\textit{u}(\textbf{q})$ (fig. \ref{distributions}.a) and $ \phi(\textbf{q})$ (fig. \ref{distributions}.b) extracted from graphene and  2D turbulence systems.

Using different distributions with the FFM, we  were able to generate surfaces that are statistically similar to those in graphene and the vorticity fields in 2D turbulence. This allowed us to investigate how different distributions influence the resulting random surfaces.

\subsubsection*{Fourier phases}

\begin{figure}[ht]
\centering
\includegraphics[scale=0.5]{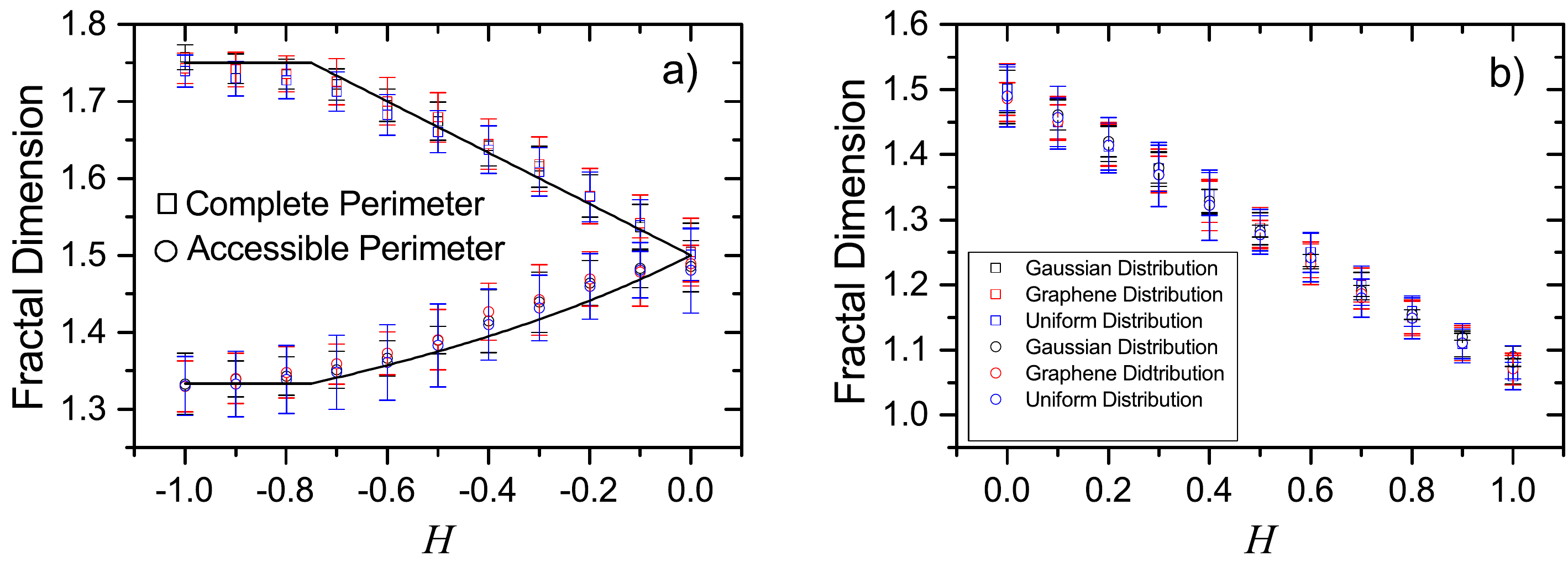}
\caption{Fractal dimension of the complete and accessible perimeters as a function of $H$, for a) $H<0$ and b) $H>0$, and different $\phi(\textbf{q})$ distributions. In a), the black lines are conjectures proposed by Schrenk K. J. \textit{et al.} \cite{Schrenk2013PercolationDisorder}. All values are averages over at least $10^{4}$ samples and error bars are defined by the variance of the distribution.}
\label{perimetersFractal}
\end{figure}

We start by showing the results obtained from using three different distributions for $ \phi(\textbf{q})$ (Gaussian, uniform, and the one found by Giordanelli \textit{et al.} in graphene) while always keeping the same Gaussian distribution for $ \textit{u}(\textbf{q})$. Applying the method described in the previous section, we obtained the dependence of the fractal dimension of the complete ($d^{com}_{f,H}$) and accessible ($d^{acc}_{f,H}$) perimeters on $H$, as illustrated in
fig. \ref{perimetersFractal}. Since exact values for the fractal dimension of those perimeters are known only for $H=-1$ and $H=0$, all other proposed analytical dependencies on $H$ are conjectures supported by numerical results \cite{Schrenk2013PercolationDisorder,Lawler2001ValuesExponents,Lawler2001ValuesExponentsb,Voss1984TheHulls,Sapoval1985ThePercolation}.
In the case of uncorrelated surfaces, $d^{com}_{f,H=-1}=7/4$ and $d^{acc}_{f,H=-1}=13/10$.
When $H$ increases from $-1$, the fractal dimension of complete and accessible perimeters start to converge.
Once the surfaces are described by a discrete Gaussian Free Field \cite{Lodhia2014FractionalSurvey} for $H=0$, the results becoming to $df^{com}_{H=0} = df^{acc}_{H=0}=3/2$.
Our results therefore point towards the absence of any dependence of $df^{com}_{H}$ and $ df^{acc}_{H}$ on the shape of the distribution of $\phi(\textbf{q})$. As shown in fig. \ref{perimetersFractal}, the $H$-dependence of $df^{com}_{H}$ and $ df^{acc}_{H}$ agrees with the conjectures made Schrenk K. J. \cite{Schrenk2013PercolationDisorder} for both, long-range correlated (fig. \ref{perimetersFractal}.a) and rough surfaces (fig. \ref{perimetersFractal}.b).

\begin{figure}[ht]
\centering
\includegraphics[scale=0.25]{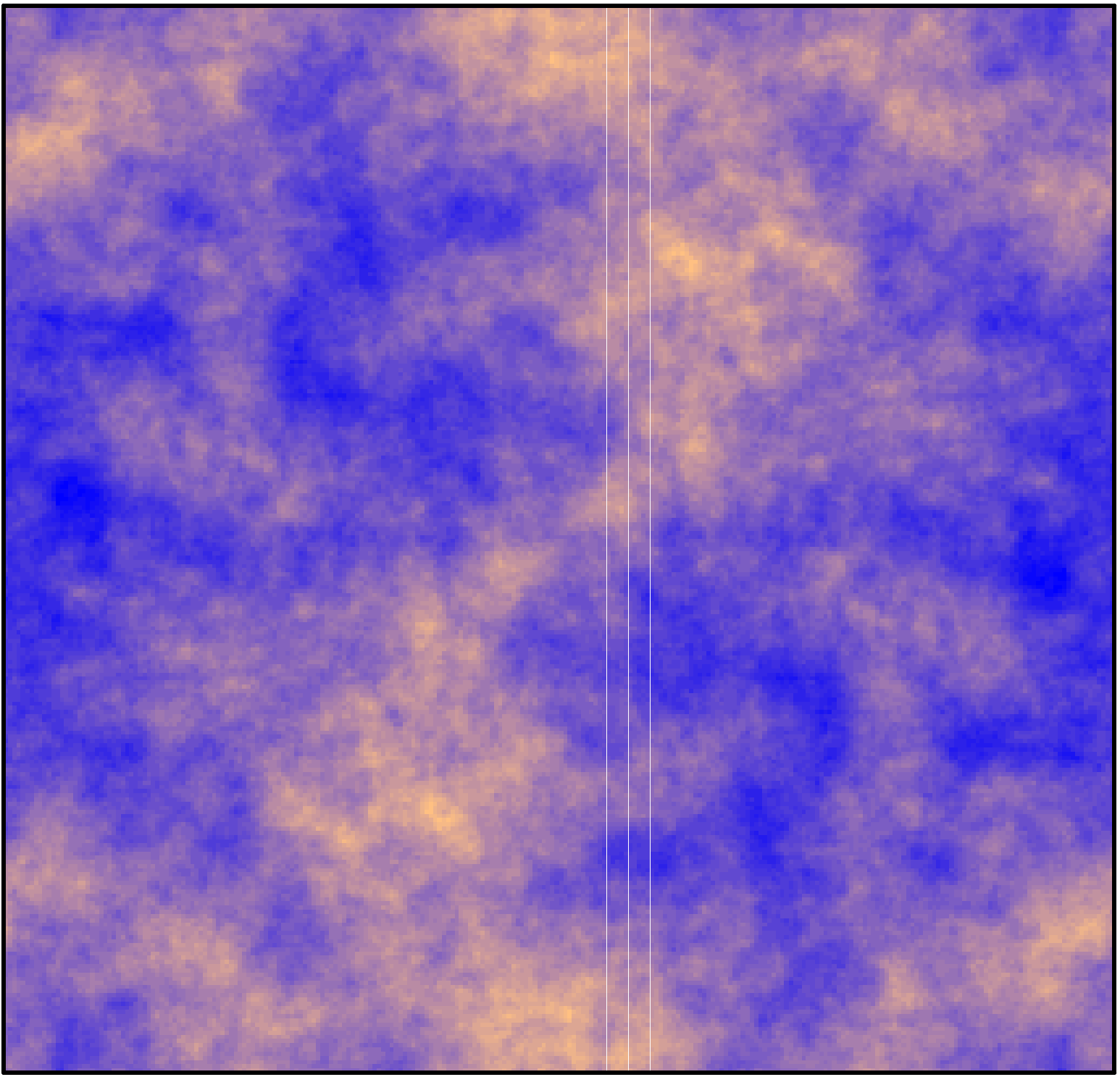}
\caption{Surface map with inversion symmetry with respect to the center. This symmetry of the surface results from the use of a Gaussian distribution  $\phi(\textbf{q})$ with a small variance $\sigma = 0.001$.} 
\label{mirrorSurface}
\end{figure}

At first glance, the influence of the Fourier phases on the random surface might not be obvious.
However, we notice that the  phase mainly influences inversion symmetries with respect to the center of the surface as shown in fig. \ref{mirrorSurface}.
In order to illustrate  the effect
we used a Gaussian distribution for $\phi(\textbf{q})$ with variance $\sigma$ close to zero. In fig. \ref{mirrorSurface} it is possible to identify the same morphological structures when the figure is rotated by an angle $\pi$. We have also found the same symmetry for different distributions of $ \textit{u}(\textbf{q})$.  This implies that, regardless of the distribution of the Fourier coefficient magnitudes, the symmetries of a random surface depends on whether the distribution of its Fourier phases is uniform or not.

\subsubsection*{Correlated phases}

\begin{figure}[ht]
\centering
\includegraphics[scale=0.5]{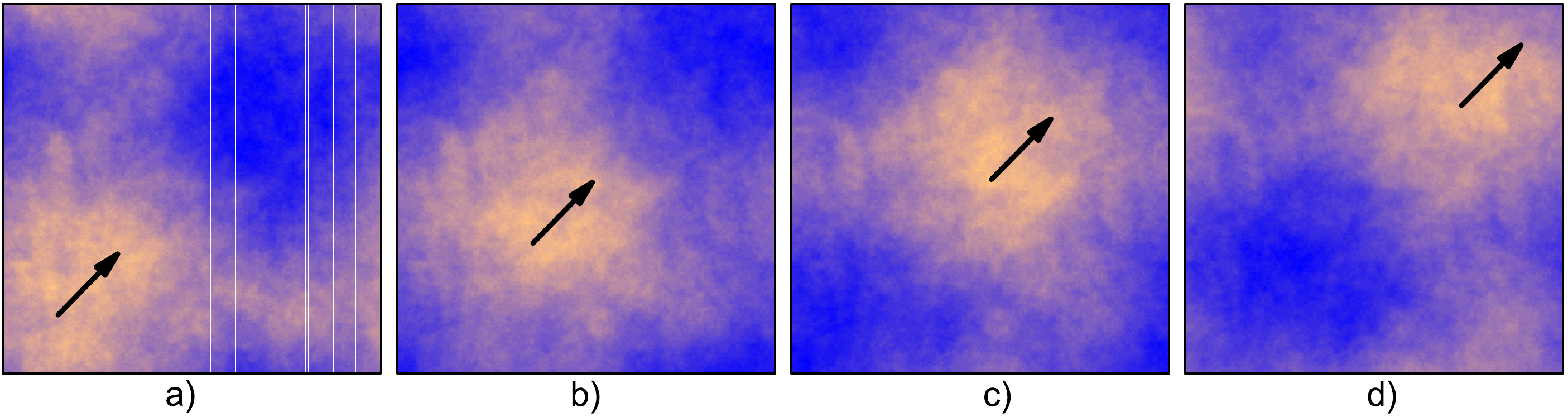}
\caption{ Maps of phase correlated surfaces. Panels a), b), c), and d) show examples of surfaces with $H=0.5$ and $H_{phase}=-0.9, -0.2, 0.1,$ and $0.4$ respectively. The arrows serve as a guide to show the linear translation of the random surface due to correlations introduced between the Fourier phases.}
\label{phaseTranslation}
\end{figure}

It turns out that the Fourier phases from the vorticity fields and graphene sheets that we analyzed are uncorrelated. 
Nevertheless, in order to understand how correlations affect random surfaces, we generated some samples with artificially correlated Fourier phases. For this purpose, we  introduced correlations in the Fourier phases  by applying the FFM twice. First we used the FFM to create a surface of correlated random phases in $\mathbf{q}$-space with Hurst exponent $H_{\varphi}$.
This surface corresponds to the phase function in eq. \ref{coefficient} with Hurst exponent $H_{\varphi}$. Applying the FFM again, we  generate Gaussian surfaces  with Hurst exponent $H$ and with the desired coefficients and correlated Fourier phases. Using always the same distributions of $\phi(\textbf{q})$ and $ \textit{u}(\textbf{q})$ and keeping fixed the value of $H$ and the seed of the random number generator, we studied the changes in the surface caused by a change in $H_{\varphi}$. We found that the correlation of Fourier phases causes a linear translation of the random surfaces (fig. \ref{phaseTranslation}). A change in $H_{\varphi}$ modifies the slope of the power spectrum (eq. \ref{power}), causing all sites of $\phi(\textbf{q})$ to shift proportionally. The random surface is affected by a linear translation because a phase shift corresponds to a translation in real space \cite{SmithStevenW.TheProcessing}.

\subsubsection*{Magnitude of the Fourier coefficients}

\begin{figure}[ht]
\centering
\includegraphics[scale=0.3]{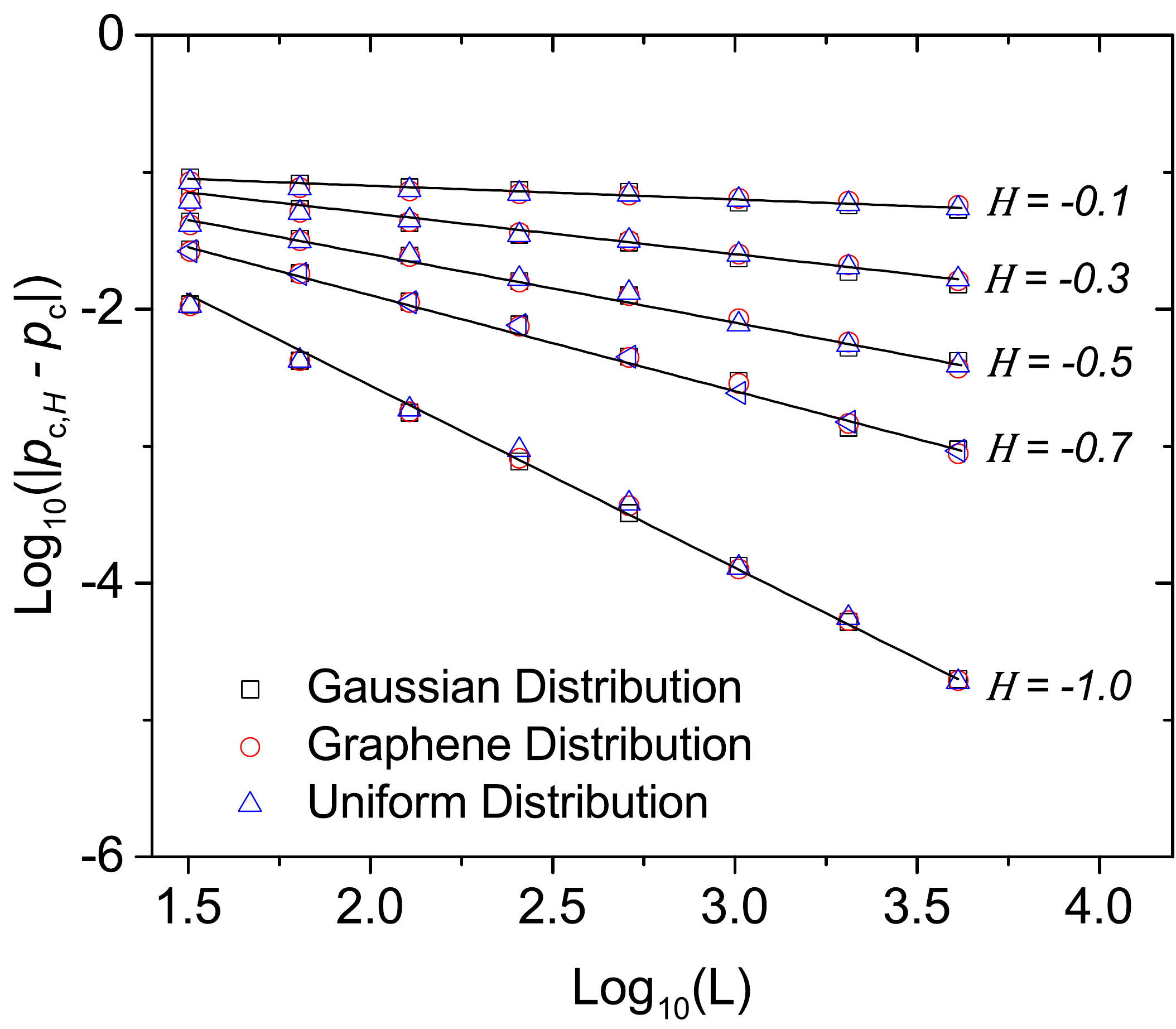}
\caption{Scale analysis of the convergence of the percolation threshold $p_{c,H}$. For the square lattice, the site percolation threshold $p_{c}$ for uncorrelated surfaces is $p_{c} \simeq 0.592746$. The black lines serve as guides to the eye with slope $H = -1/\nu_{H}$ \cite{Newman2000EfficientPercolation,Newman2001FastPercolation,Schrenk2013PercolationDisorder}.}
\label{nu}
\end{figure}

We generated sets of random surfaces, each with $\phi(\textbf{q})$ uniformly distributed but with a different distribution of $\textit{u}(\textbf{q})$: Gaussian, uniform and the distribution found by Giordanelli \textit{et al.} in graphene. We then determined the average values of two critical exponents of percolation corresponding to each set of surfaces with a different $\textit{u}$-distributions. We first considered the $H$-dependence of the correlation length critical exponent $\nu_{H}$   for $ -1 \leq H \leq 0  $. It is well established that the critical point $p_c \simeq 0.592746$ \cite{Dietrich1985IntroductionTheory,Weinrib1984Long-rangePercolation,Schmittbuhl1993PercolationSurfaces,Weinrib1983CriticalDisorder} is the infinite system size limit of the percolation threshold $p_c(H,L)$ which is $H$-dependent for finite system sizes, $L$. Furthermore, the expected scaling behavior \cite{Newman2000EfficientPercolation,Newman2001FastPercolation,Schrenk2013PercolationDisorder} is   
\begin{equation}\label{eq:scaling}
|p_{c}(H,L) - p_{c}| \sim L^{-1/\nu_{H}},
\end{equation}
with $\nu_{H} = -1/H$ \cite{Schrenk2013PercolationDisorder,Manna2012AboutPercolation,Schrenk2013StackedProperties}. Our numerical results in fig. \ref{nu} not only confirm that the scaling relation in eq. \ref{eq:scaling} is respected no matter which one of the three $u$-distributions we use but also that the value of $p_c$ remains unchanged.

A consequence of the scaling relation in eq. \ref{eq:scaling} is that in the asymptotic limit it is sufficient to compute the critical exponents at the percolation threshold, $p_c$. 

\begin{figure}[ht]
\centering
\includegraphics[scale=0.3]{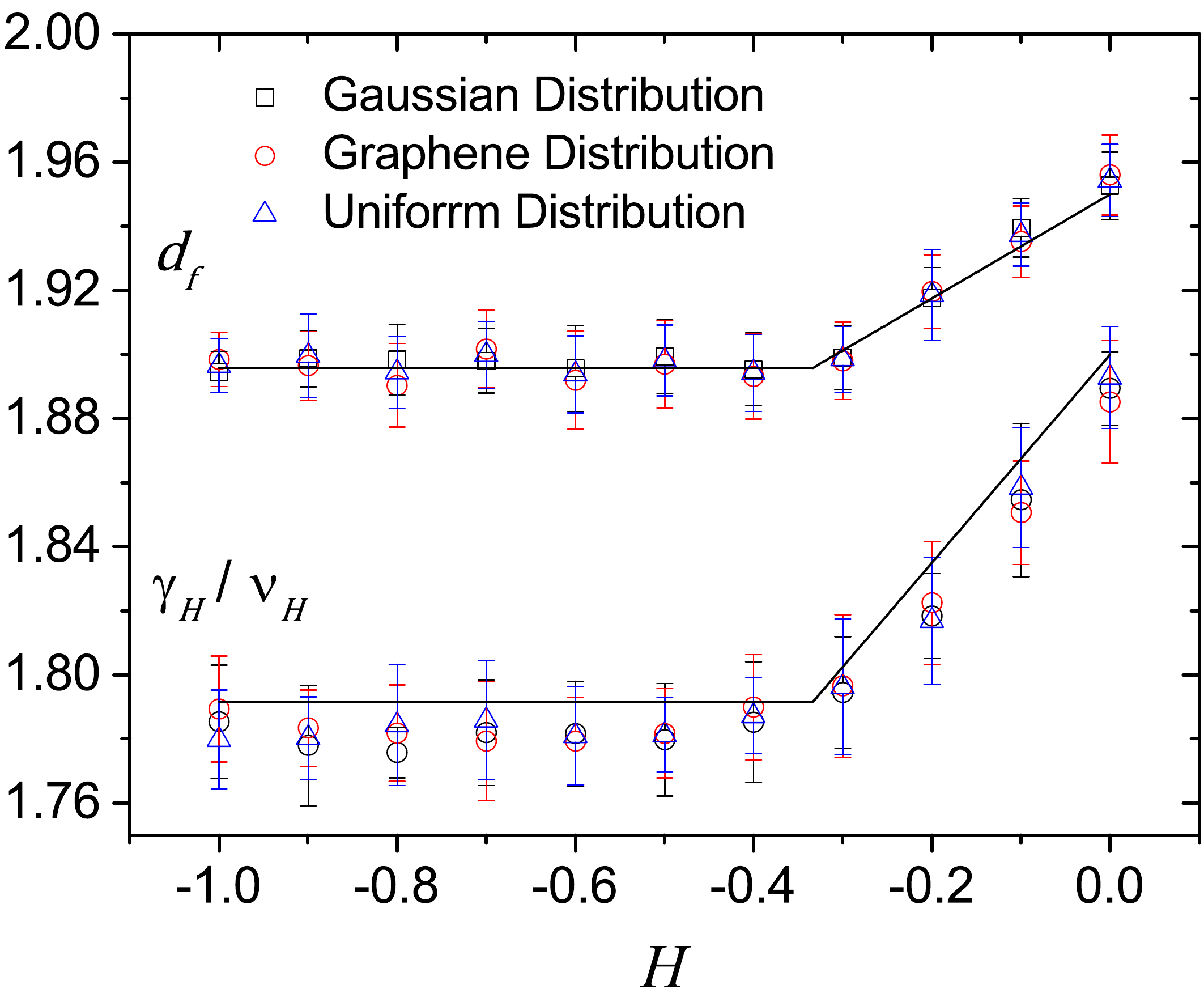}
\caption{Fractal dimension $d_{f}$ of the percolation cluster  and critical exponent ratio $\gamma_{H}/\nu_{H}$ as a function of the Hurst exponent $H$ for surfaces with different  distributions of $ \textit{u}(\textbf{q})$. The black lines are conjectures proposed by Schrenk K. J. \textit{et al.} \cite{Schrenk2013PercolationDisorder} based on the hyperscaling relation \cite{Dietrich1985IntroductionTheory}. All values are averages over at least $10^{4}$ realizations	 and error bars are defined as the variance of the distribution of their values.}
\label{clusterFractalDimension}
\end{figure}

At this critical point, the percolation cluster is a fractal with fractal dimension $d_{f}$.
The occupancy, $M_{max}$, which is the number of sites that belong to the percolation cluster, scales with lattice size $L$ as,
\begin{equation}
M_{max} \sim L^{d_{f}}.
\label{clusterDf}
\end{equation}
Using eq. \ref{clusterDf} we recovered numerically the value of the fractal dimension as a function of the Hurst exponent. In fig. \ref{clusterFractalDimension} our results show that the value of the fractal dimension of the percolation cluster remains the same for all three distributions of $\textit{u}(\textbf{q})$.

We also checked the $H$-dependence of the susceptibility critical exponent $\gamma$ by considering the scaling behavior of $m_{2}$, the second moment of the distribution of the cluster sizes at $p_{c}$ defined as \cite{Dietrich1985IntroductionTheory}
\begin{equation}
m_{2} = \sum_{k} \frac{M_{k}^{2}}{N} - \frac{M_{max}^{2}}{N}.
\end{equation}
Here, the sum goes over all clusters, where $ M_{k} $ is the mass of cluster $k$, and we use the fact that the following scaling behavior holds at \cite{Dietrich1985IntroductionTheory}:
\begin{equation}
m_{2}\sim L^{\gamma_{H}/\nu_{H}}.
\end{equation}

For uncorrelated percolation $(H=-1)$, $\gamma_{H=-1} =43/18$, $ \nu_{H=-1}=4/3$ such that $d_{f}=91/48$  and $\gamma_{H=-1}/\nu_{H=-1}=43/24$ \cite{Dietrich1985IntroductionTheory}. Fig. \ref{clusterFractalDimension} shows the dependence on $H\in[-1,0]$ of both critical exponents, the fractal dimension of the percolation cluster and the exponent ratio $\gamma/\nu$, for different distributions of $u(\textbf{q})$.

In conclusion, our results suggest that both exponents, $d_f$ and $\gamma/\nu$, are independent of the distribution of $u(\textbf{q})$. In fact, the only change that we identified was in the height values $h(\mathbf{x})$ of the random surfaces.

\section*{Conclusions}

We considered two concrete examples of random surfaces, namely, the vorticity field of turbulent systems in two dimensions and rough graphene sheets. We investigated how these random surfaces and in particular the critical exponents are influenced by the presence of phase correlations and by changes in the distribution of the Fourier coefficient magnitudes and Fourier phases. Our results show that the Fourier phases distribution of the vorticity field and graphene sheets, within error bars, lead to the same value for the fractal dimension of the complete and accessible perimeters. We also showed that any phase correlation in Fourier space leads to a translation of the random surfaces, and that they do not have any influence on their statistical properties. For different  distributions of magnitude of Fourier coefficients our results suggest there is no $H$ dependence of the fractal dimension of the percolation cluster and susceptibility exponent. In addition, we recovered for the critical exponents the same $H$-dependence as conjectured by Schrenk K. J. \cite{Schrenk2013PercolationDisorder}. Although we have only considered three examples of Fourier coefficient distributions, we do not expect different results for any other distribution with finite variance.

\bibliography{Mendeley}
\bibliographystyle{unsrt}

\section*{Acknowledgements}

We acknowledge the financial support from European Research Council (ERC) Advanced Grant 319968 FlowCCS, the ETH Risk Center, the Brazilian INCT-SC, and Minist\'erio da Educa\c{c}\~ao do Brasil (Funda\c{c}\~ao CAPES). 
We also thank M. Mendoza and I. Giordanelli for providing surface data.

\section*{Author contributions statement}

C. P. de Castro, M. Lukovi\'c, R. F. S. Andrade and H. J. Herrmann conceived the research, C. P. de Castro conducted the numerical simulations,. All authors contributed to the writing of the manuscript.

\section*{Additional information}

\textbf{Accession codes} (where applicable); 
\textbf{Competing financial interests} The authors declare no competing financial interests. 


\end{document}